\documentclass[reprint,superscriptaddress,amsmath,amsfonts,amssymb,aps,prl,showkeys]{revtex4-1}
\usepackage{bm}
\usepackage{color}
\usepackage{graphicx}
\usepackage[normalem]{ulem}
\usepackage[makeroom]{cancel}

\usepackage{hyperref}
\hypersetup{
  breaklinks = {true},
  citecolor = {blue},
  colorlinks = {true},
  linkcolor = {red},
  urlcolor = {blue},
}

\graphicspath{{../Figs/}}

\begin{document}

\title{Converting Non-Equilibrium Charge Density into Spin Current}

\author{Marc Vila}
\email{marcvila@berkeley.edu}
\affiliation{Department of Physics, University of California, Berkeley, California 94720, USA}
\affiliation{Materials Sciences Division, Lawrence Berkeley National Laboratory, Berkeley, California 94720, USA}

\author{Joel E. Moore}
\affiliation{Department of Physics, University of California, Berkeley, California 94720, USA}
\affiliation{Materials Sciences Division, Lawrence Berkeley National Laboratory, Berkeley, California 94720, USA}	

\date{\today}

\begin{abstract}
The interconversion between charge and spin degrees of freedom is of both fundamental and technological relevance in spintronics. While a non-equilibrium spin density and a charge current are related by the well known Rashba-Edelstein effect, here we theoretically model the generation of a time-dependent spin current due to a periodic modulation of the charge density, for example by a gate. By using the Boltzmann transport equation, we show that when the chemical potential is varied, a spin current is generated in the time scale it takes for the system to re-equilibrate in the new chemical potential. The effect is ubiquitous in many systems with spin-momentum locking; we compute the strength of the effect in four examples and propose a simple device scheme to measure the spin accumulation resulting from such time-dependent spin currents. Our findings test fundamental theoretical questions about charge-to-spin conversion mechanisms and provide an all-electrical way to generate spin currents without the need for charge currents, magnetic materials or optical methods.
\end{abstract}

\maketitle

%

\textit{Introduction.---} Pure spin currents, which transport the electron spin degree of freedom without charge, are crucial in many types of spintronic devices and phenomena, and hence simple and efficient ways to create and detect them are of fundamental importance. For instance, spin currents are the building block of spin logic devices \cite{Datta1990, Koo2009, Chuang2015} or spin torque-based magnetic memories \cite{Gambardella2011, Chernyshov2009, Miron2010} and are also being used to study fundamental properties such as spin relaxation times \cite{Zutic2004, Fabian2007}. Different widely employed methods exist to generate spin currents in the solid state. Optical spin orientation has been widely used in semicondcutors, where selection rules allow for selectively exciting a specific spin polarization \cite{Parson1969, Zutic2004}. Spin pumping \cite{Tserkovnyak2002, Costache2006, Cheng2014} can excite precession of the magnetization of ferromagnets to inject spins in adjacent materials. However, perhaps one of the most attractive ways to achieve spin currents is by means of charge-to-spin interconversion mechanisms. This enables all-electrical systems without the need to integrate optics or magnetic materials, thus decreasing the complexity and fabrication processes of devices.

The interconversion between charge and spin dates back more than fifty years ago \cite{Dyakonov1971, Dyakonov1971b} and has been a major topic in spintronics over the years \cite{Sinova2015}. The main mechanisms include the spin Hall effect \cite{Hirsch1999, Zhang2000, Sinova2004, Kato2004, Wunderlich2005, Sih2005} and the Rashba-Edelstein effect \cite{Edelstein1990, Kato2004PRL, Silov2004, Sih2005}. The former converts charge current into spin current, while the latter converts charge current into non-equilibrium spin density, and both have their reciprocal effects (i.e. spin-to-charge conversion) \cite{Valenzuela2006, Ganichev2002, Rojas2013, Shen2014}. Nevertheless, the electrical generation of strong spin currents without charge currents has eluded the family of charge-to-spin interconversion phenomena so far. Such a mechanism could create spins without the Joule heating inherent from electrical currents, thereby enabling energy-efficient magnetization switching and spintronic devices \cite{Razavi2017, Li2018, Liu2021}.



In this work, we propose a mechanism for creating time-dependent spin currents from a non-equilibrium charge density. We use the Boltzmann transport equation for the spin current to show that when the chemical potential is varied, a spin current is generated in the time scale it takes for the system to re-equilibrate in the new chemical potential. We illustrate such an effect in different paradigmatic Rashba models as well as a model of chiral tellurium. We also suggest a simple setup to probe this effect in which the resulting non-equilibrium spin accumulation follows the same spatial profile as that of the spin Hall effect, thus allowing for common optical and electrical detection schemes. Our results broadens the family of charge-to-spin interconversion processes and opens new avenues for more energy-efficient spintronic devices.

\begin{figure}[bt]
    \centering
    \includegraphics[scale=0.48]{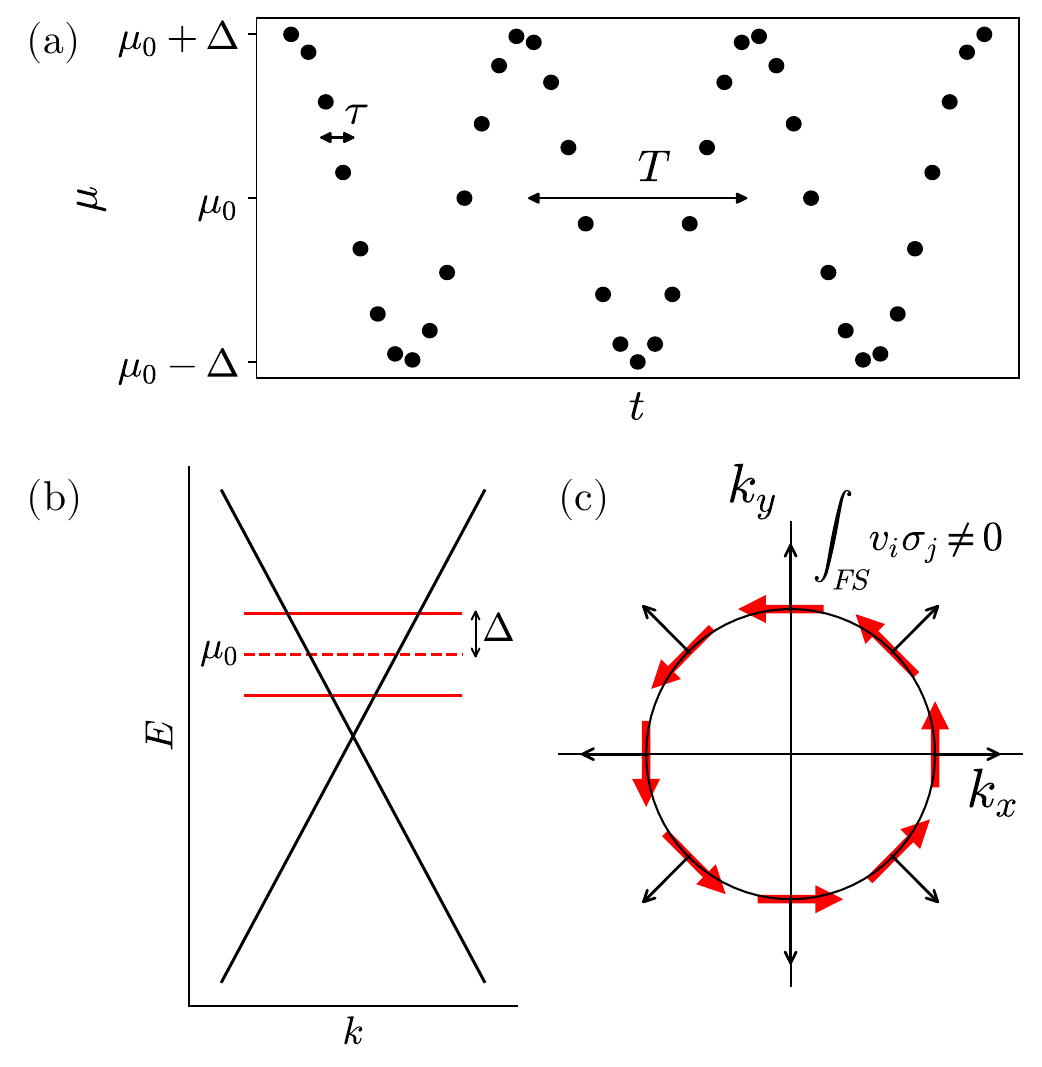}
    \caption{(a) Variation of the chemical potential with time. Electrons take a time $\tau$ to reach equilibrium in the new chemical potential. (b) Schematics of the variation of the chemical potential in the band structure, where $\mu$ ranges from $\mu_0 - \Delta$ to $\mu_0 + \Delta$. (c) Fermi surface of single-band Rashba model, with red (black) arrows denoting the spin texture (band velocity). In such a system, the product of the spin texture and velocity integrated throughout the Fermi surface is nonzero.}
    \label{fig_fig1}
\end{figure}

\textit{Boltzmann equation.---} In the framework of the Boltzmann transport equation, one can express the spin current density in an homogeneous system as \cite{Culcer2004, Shi2006}
\begin{equation}\label{eq_Js}
\textbf{J}_s^i  = \frac{\hbar}{2} \frac{1}{4 \pi^2} \int_{BZ} f(\textbf{k},t) \textbf{v}(\textbf{k}) \langle \sigma_i (\textbf{k}) \rangle   \text{d} \textbf{k},
\end{equation}
where $\hbar$ is the reduced Planck constant, $\textbf{v}$ is the electron velocity, $\sigma_i$ is the spin Pauli matrix with $i=x,y,z$ and $f$ the deviation of the electron distribution function from its equilibrium one, $f_0$. This definition automatically satisfies the expectation that a measurable transport current of spin requires that the system be out of equilibrium. Commonly, Eq. \eqref{eq_Js} is used for the Rashba-Edelstein effect \cite{Manchon2008, Manchon2009, Gambardella2011}, in which a constant electric field in the transport direction shifts the Fermi surface in momentum space, leading to unequal population of $+\textbf{k}$ and $-\textbf{k}$ states and thus finite $f(\textbf{k},t)$. Here, we propose a scenario where the out-of-equilibrium condition is reached in the time domain while maintaining the same number of $+\textbf{k}$ and $-\textbf{k}$ states. For that, we consider the case of an alternating electric field (generated by e.g. a gate voltage) in a two-dimensional system that periodically modulates the chemical potential $\mu$ by constantly changing the electron density, as illustrated in Figs. \ref{fig_fig1}(a) and (b). In this way, $f_0$ is a time-dependent Fermi-Dirac distribution function:
\begin{equation}
f_0 (t) = \frac{1}{1 + e^{\beta (\varepsilon(\textbf{k}) - \mu(t))}},
\end{equation}
with $\mu (t) = \mu_0 + \delta\mu(t) = \mu_0 + \Delta \cos(\omega t)$. Here, $\beta$ is the inverse temperature, $\mu_0$ is the chemical potential in absence of any perturbation and $\Delta$ quantifies the variation of the chemical potential with respect to the electric field (with frequency $\omega$). Upon the application of a gate voltage, the system reacts to change its carrier density and therefore its chemical potential.  While there are situations in semiconductors at moderate temperatures in which the Coulomb interaction between electrons can act to reduce spin currents via a kind of drag effect~\cite{damicovignale,orenstein}, we follow most work in ignoring such interactions, and we also assume that the changing gate voltage does not modify the electronic states, only their occupations.

Although fast, this change is not instantaneous in real materials, and the distribution takes some time $\tau$ to reach a new equilibrium (see Fig. \ref{fig_fig1}(a)). Therefore, at a given time $t$, we can formulate an out-of-equilibrium distribution function, $f_1$.
To model $f_1$, and hence $f$, we resort to the Boltzmann equation within the relaxation time approximation: 
\begin{equation}\label{eq_Boltzmann}
\frac{\partial f_1}{\partial t} = -\frac{f}{\tau} = - \frac{f_1 - f_0}{\tau}.
\end{equation}
Since it is reasonable to assume that $\delta \mu$ is small, we linearize $f_0$ with respect to $\Delta$ to solve Eq. \eqref{eq_Boltzmann} analytically. This results in
\begin{equation}\label{eq_f0}
f_0 (t) = f_0^{\Delta=0} + \delta f_0^{\Delta=0} \Delta \cos(\omega t) ,
\end{equation}
with $f_0^{\Delta=0}$ being the equilibrium distribution in the absence of an electric field (i.e. time-independent) and $\delta f_0^{\Delta=0} = \frac{\partial f_0^{\Delta=0}}{\partial \varepsilon}$. Eq. \eqref{eq_f0} tells us that for energies far from the Fermi level, the occupation number will not vary with time. Only at energies within $\delta f_0^{\Delta=0}$ will the carrier density vary at a frequency $\omega$ and this change will be proportional to the variation of chemical potential $\Delta$. Since Eq. \eqref{eq_Boltzmann} is an inhomogeneous linear differential equation, the general solution for arbitrary initial conditions is obtained by adding any solution of the homogeneous equation, which here is just a decaying transient $\sim e^{-t/\tau}$, to a single solution of the inhomogeneous equation, which we now find for the simple initial condition  $f_1(t=0)=f_0(t=0)$.  Plugging Eq. \eqref{eq_f0} into Eq. \eqref{eq_Boltzmann}, we obtain:
\begin{align}\label{eq_f1}
f_1 &= [ \delta f_0^{\Delta=0} \Delta  (\omega \tau)^2 e^{-t/\tau} +  f_0^{\Delta=0} +  f_0^{\Delta=0}(\omega \tau)^2 \nonumber \\  
&+ \delta f_0^{\Delta=0} \Delta \cos(\omega t) + \delta f_0^{\Delta=0} \omega \tau \Delta \sin(\omega t) ] \frac{1}{1+(\omega \tau)^2}.
\end{align}
Assuming that $\tau$ is the shortest timescale of the problem, we take the limit of $\omega \tau \ll 1$ in Eq. \eqref{eq_f1}, subtract Eq. \eqref{eq_f0} and keep only the leading terms in $\omega \tau$, which happens to be linear order. In this way, we obtain $f$ as
\begin{equation}
f = \delta f_0^{\Delta=0} \omega \tau \Delta \sin(\omega t).
\end{equation}
In the low-temperature limit, we can recast Eq. \eqref{eq_Js} as a Fermi surface integral:
\begin{equation}\label{eq_Js2}
\textbf{J}_s^i = \frac{\hbar}{2} \frac{\omega \tau \Delta \sin(\omega t)}{4 \pi^2 \hbar} \int_{FS}  \frac{\textbf{v}(\textbf{k}) \langle \sigma_i (\textbf{k}) \rangle}{|\textbf{v}(\textbf{k})|}   \text{d} \textbf{S}_F. 
\end{equation}
The above expression resembles the one obtained for the Rashba-Edelstein effect \cite{Gambardella2011} with the exception that such an effect computes the non-equilibrium spin {\it density} whereas Eq. \eqref{eq_Js2} is for the spin {\it current}. Furthermore, in our case, the out-of-equilibium distribution function $f(\textbf{k},t)$ arises due to a time-dependence of the electron density (reflected in a change of the chemical potential) while in the Rashba-Edelstein effect $f(\textbf{k})$ is a consequence of the applied bias current. In this way, our proposed mechanism converts non-equilibrium charge density into spin current, as opposed to the Rashba-Edelstein effect where charge current is converted into non-equilibrium spin density.  

\textit{Rashba models.---} So far we have obtained the distribution function of a system with periodic oscillations of the chemical potential, but this alone does not guarantee a finite spin current given by Eq. \eqref{eq_Js2}. To maximize the integral of the term $\textbf{v}(\textbf{k}) \langle \sigma_i (\textbf{k}) \rangle$,  a system with spin-momentum locking is ideal, as both the spin expectation value and the band velocity are odd functions of momentum, as shown in Fig. \ref{fig_fig1}(c). Materials with Rashba spin-orbit coupling \cite{Manchon2015}, such as inversion-asymmetric semiconductors, van der Waals heterostructures or surface states of topological insulators such as Bi$_2$Te$_3$ present that kind of spin texture. Consequently, we compute the spin current for three representative Rashba models corresponding to 2DEGs, graphene and TI surface states. The effective Hamiltonians read
\begin{align}
&H_{\text{2DEG}} = \frac{k^2\hbar^2}{2 m^*} + \alpha(\bm{\sigma} \times \textbf{k})\cdot \hat{\textbf{z}} \\
&H_{\text{Gr}} = \hbar v_F^\text{Gr} (\tau \mu_x k_x + \mu_y k_y) + \lambda_\text{Gr}(\tau \mu_x \sigma_y - \mu_y \sigma_x)\\
&H_{\text{TI}} = \lambda(\bm{\sigma} \times \textbf{k})\cdot \hat{\textbf{z}} 
\end{align}
where $m^*$ is the effective mass, $v_F^\text{Gr}$  and $\lambda=\hbar v_F^\text{TI}$ are the Fermi velocities of graphene and TI surface states, respectively, $\tau=\pm 1$ is graphene's valley index, $\mu_{x,y}$ are the Pauli matrices acting on graphene's pseudospin and $\alpha$ and $\lambda_\text{Gr}$ are the Rashba spin-orbit strength of the 2DEG and graphene, respectively. It is worth noting that the energy dispersion of TI surface states, $\lambda$, is also related to the spin-momentum locking strength. Because the energy dispersions, band velocities and spin textures can all be obtained analytically, we solve Eq. \eqref{eq_Js2} analytically for each model. In the limit of $|\mu_0| > |\Delta|$ (i.e. only states above or below the band gap/Dirac point contribute), the spin currents read:
\begin{align}
&{J}_{s,y}^{x,\text{2DEG}} = - J_0 \frac{2 m^* \alpha}{\hbar^2} \label{eq_Jsmodels} \\
&{J}_{s,y}^{x,\text{Gr}} = J_0 \frac{4 \lambda_\text{Gr}}{\hbar v_F^\text{Gr}} \frac{\left(\mu_0 + \Delta \cos(\omega t)\right)^2}{\left(\mu_0 + \Delta \cos(\omega t)\right)^2-\lambda_\text{Gr}^2} \approx J_0 \frac{4 \lambda_\text{Gr}}{\hbar v_F^\text{Gr}} \label{eq_Jsmodels2} \\ 
&{J}_{s,y}^{x,\text{TI}} = J_0 \frac{\mu_0 + \Delta \cos(\omega t)}{\lambda} \label{eq_Jsmodels3},
\end{align}

with $J_0 = \frac{\hbar}{2} \frac{\omega \tau \Delta \sin(\omega t)}{4 \pi \hbar}$ and summation over bands and valleys are performed, where necessary. A few comments are in order here. These expressions are valid for bands above and below the band gap/Dirac point, although with opposite sign for the 2DEG and graphene cases. Also, ${J}_{s,y}^{x} = -{J}_{s,x}^{y}$, ${J}_{s,x}^{x} = {J}_{s,y}^{y} = 0$. We restrict these expressions to the cartesian directions $x$ and $y$, but we note that the spin current can be written along any direction in the $xy$ plane, with the direction of the spin just being perpendicular to that. Although these properties are qualitatively similar to all models, there are important quantitative differences. For instance, the spin current in 2DEGs and graphene is independent on the original chemical potential $\mu_0$ (for graphene, $\lambda_\text{Gr} << \mu_0, \Delta$), while this is not the case for TIs. This suggests that doping the TI surface states, which commonly happens and it is usually not preferred \cite{Walsh2018, Sakamoto2021}, is here advantageous for generating larger currents. Additionally, we observe that in all cases, reducing the band dispersion (either by increasing $m^*$ or decreasing $v_F^\text{Gr}$ or $\lambda$), increases the spin current. This contrasts with typical expressions for charge current, where the current is proportional to the Fermi velocity \cite{Gambardella2011}. 

\begin{figure}[bt]
    \centering
    \includegraphics[scale=0.5]{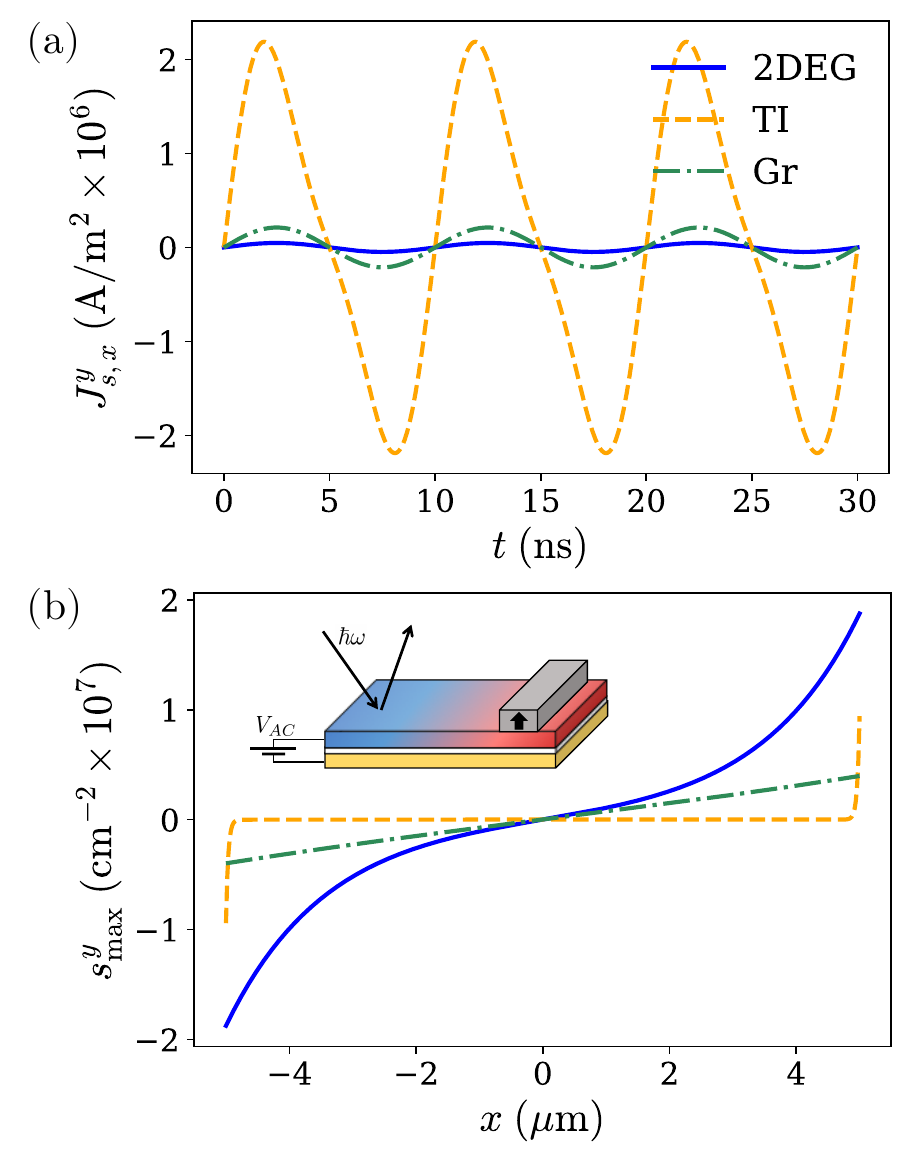}
    \caption{(a) Spin current density for three representative Rashba models computed from Eqs. \eqref{eq_Jsmodels}-\eqref{eq_Jsmodels3}. (b) Spatial profile of the spin accumulation (Eq. \eqref{eq_S}) for a time in which the spin is maximum. Inset: schematic of the proposed device to generate and measure the time-dependent spin current using optical probes or magnetic contacts. The gold and white layers are the back gate and insulator, respectively, while the top layer is the spin-orbit material showing spin accumulation (red and blue colors).}
    \label{fig_fig2}
\end{figure}

To better visualize the magnitude and behavior of the expressions above, we plot the three spin currents in Fig. \eqref{fig_fig2}(a) for representative values of parameters \cite{Suppmat} and taking a conservative value of $\tau=10$ fs. We multiply the spin currents by a factor $2e/(\hbar t)$ to convert the units into A/m$^2$, with $t$ being the effective thickness of the material ($t_\text{2DEG}=2$ nm \cite{Kong2004, Ohta2018}, $t_\text{Gr} = 0.35$ nm \cite{Neto2009}, $t_\text{TI} = 3$ nm \cite{Wu2013, Jash2021}). Amongst the three models, the TI surface states is the system with the highest spin current density, perhaps unsurprisingly since they  possess spin-orbit coupling that is large enough to induce a phase transition from a trivial phase. Graphene and 2DEGs show comparable magnitudes, with spin currents in graphene being slightly larger due to our choice of parameters, noting that the opposite could be true for larger effective masses or smaller spin-orbit in graphene.

The region where the spin current is generated and how spins propagate will depend on the device setup. We consider a simple device for convenience, consisting of a back gate below our two-dimensional system in which an AC gate voltage is used to tune the chemical potential (see Fig. \ref{fig_fig2}(b)). In this way, the spin current will be generated equally in the whole sample and will propagate radially from each point. In this configuration, one can employ the spin diffusion equations to calculate the spatial profile of the spin density carried by the spin current. The spin diffusive equation for diffusive transport reads \cite{Garcia2018, Vila2020, Vila2021}:
\begin{align}\label{eq_Bloch}
    \frac{\partial s^i}{\partial t} + \nabla \cdot \textbf{j}_s^i = -e\frac{s^i}{\tau_s},
\end{align}
where $s^i$ is the spin accumulation or density of spins with polarization along the $i$ direction and $\tau_s$ is the spin relaxation time. The spin current and density are also related by
\begin{align}\label{eq_Js_condition}
    \textbf{j}_s^i = -e D_s \nabla s^i + \textbf{J}_s^i,
\end{align}
with $D_s$ being the spin diffusion coefficient (which many times is equated to the charge diffusion coefficient) and $\textbf{J}_s^i$ is the spin current derived in this work which takes the role of a source term. If $\tau_s \ll 2\pi/\omega$, which is a good approximation in many situations, the spins relax faster than their time variation from $\textbf{J}_s^i$, and it is possible to solve Eq. \eqref{eq_Bloch} in the steady state. The boundary condition to solve the above equations is the fact that the spin current vanishes at the edge of the material, $\textbf{j}_s^i = 0$. Therefore,
\begin{align}\label{eq_boundary_condition}
    \nabla s^i |_{x=0} = \nabla s^i |_{x=L} = \frac{\textbf{J}_s^i}{e D_s}.
\end{align}
Using Eqs. \eqref{eq_Bloch}-\eqref{eq_boundary_condition}, and defining the spin relaxation length as $\lambda_s = \sqrt{D_s \tau_s}$, the spin accumulation along the $x$ direction is given by
\begin{align}\label{eq_S}
    s^y(x,t)=  -J_{s,x}^y(t) \frac{\lambda_s \sinh(\frac{L-2x}{2\lambda_s})}{e D_s \cosh(\frac{L}{2\lambda_s})}.
\end{align}
We take values of $D_s=0.005,0.05,0.015$ m$^2$/s  \cite{Bibes2012, Kuczmik2017, Raes2016, Zhou2020} and $\tau_s=500,1000,0.08$ ps \cite{Kuczmik2017, Raes2016, Liu2013} for the 2DEG, graphene and TI models, respectively, and plot the resulting spin accumulation in Fig. \eqref{fig_fig2}(b) for a system size of $L=10$ $\mu \text{m}$. We now observe that even though the 2DEG was the system with smaller spin current, it is the one with larger spin accumulation both at the edge and across the sample. This is due to a relatively large $\tau_s$ while having a rather small $D_s$. In contrast, spins in the TI surface states accumulate only at the edge and relax very fast due to a very small $\tau_s$ and $\lambda_s$.

An important aspect of Eq. \eqref{eq_S} is that it has the same functional form that the spin accumulation generated by the spin Hall effect \cite{Cornelissen2016, Stamm2017, Vila2021}. The only difference lies in the prefactor $J_{s,x}^y$, which here takes the form of Eqs. \eqref{eq_Jsmodels}-\eqref{eq_Jsmodels3} and in the spin Hall effect it is the product of the charge current and the spin Hall angle. This suggests that the same techniques employed to detect the spin Hall effect could be used here to measure our proposed spin current generation. This includes optical techniques such as the Kerr effect \cite{Kato2004, Stamm2017} and NV-center probes \cite{Wang2022, Xu2023}, or electrical methods using magnetic materials to detect nonlocal voltages or ferromagnetic resonances \cite{Costache2006, Sinova2015}.

\textit{Te model.---} The calculations performed above are based on model systems which are enough to illustrate the proposed charge-to-spin-current conversion. Nevertheless, we now apply our theory to elemental tellurium, which has recently attracted a lot of attention because of its spintronic properties tied to its crystal chirality \cite{Calavalle2022, Roy2022}. Chiral materials are known to host so-called Kramers-Weyl fermions displaying radial spin-momentum locking \cite{Chang2018, Lin2022, Gosalbez2023}, and such spin texture has recently been confirmed experimentally in tellurium \cite{Sakano2020, Gatti2020}. Importantly for our purposes, the Fermi level of hole-doped chiral tellurium nanowires was shown to be tuned by a gate voltage \cite{Calavalle2022}, which makes it an ideal candidate material to test our theoretical proposal.

\begin{figure}[bt]
    \centering
    \includegraphics[scale=0.5]{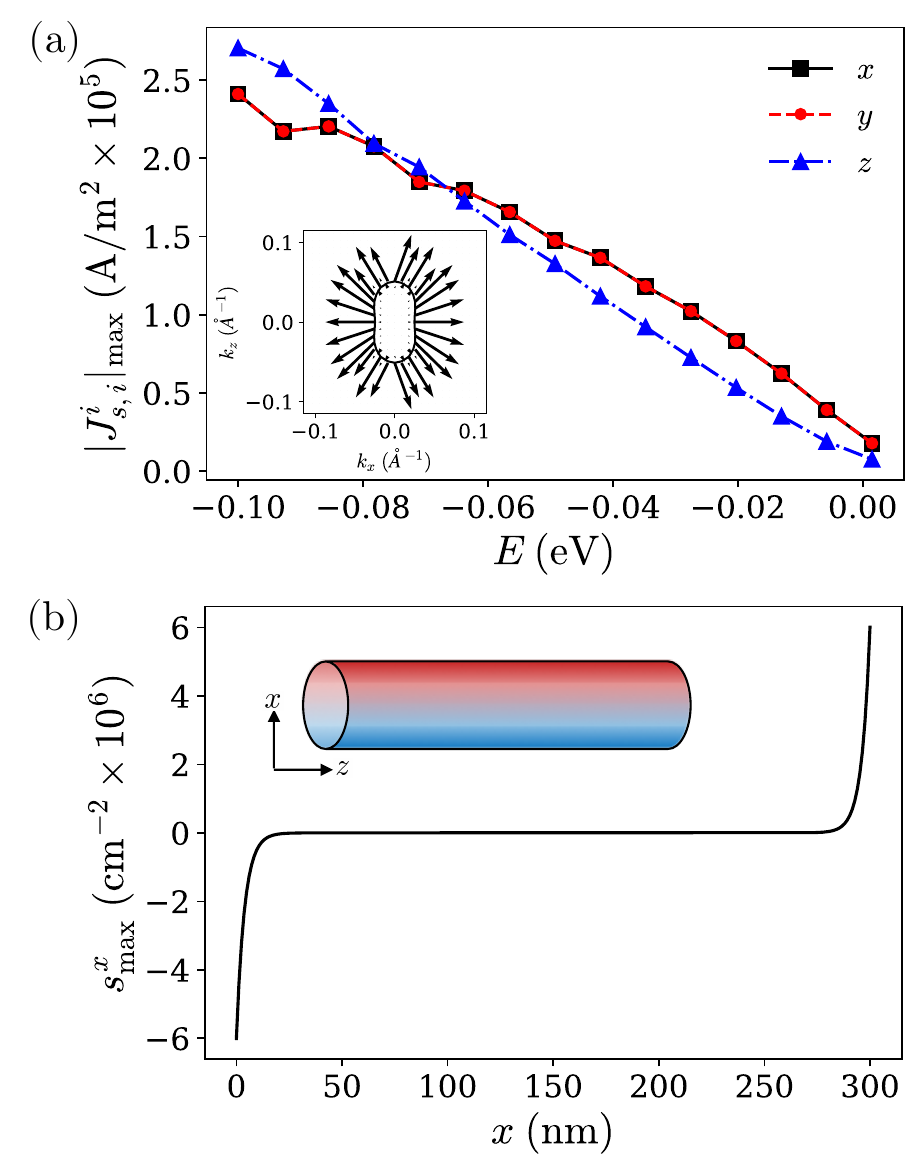}
\caption{(a) Maximum spin current density in the Te model as a function of energy for states in the topmost valence band (the second band appears at an energy $E \approx -126$ meV). Inset: Fermi surface at $E=-0.020$ eV with the arrows representing the magnitude and direction of the spin texture in the $xz$ plane. (b) Spatial profile of the spin accumulation (Eq. \eqref{eq_S}) for a time in which the spin is maximum, at $E=-0.020$ eV. Inset: schematics of the spin accumulation in the surface of the nanowire.}
    \label{fig_fig3}
\end{figure}

For a given crystal chirality, the hole physics in Te can be well captured by a 2-band model around the H point in the Brillouin zone \cite{Calavalle2022, Shalygin2012, Glazov2022}:
\begin{align}
    H_\text{Te} =& - \delta - A k_z^2 - B (k_x^2 + k_y^2) \nonumber \\
    &+  \beta k_z \sigma_z + \frac{\delta}{\sqrt{k_x^2 + k_y^2}}(k_x \sigma_x + k_y \sigma_y).
\end{align}
Here, $A$ and $B$ characterize the band dispersion and $\beta$ and $\delta$ denote the spin-orbit splitting; both quantities are isotropic within the $xy$ plane and anisotropic with respect to the $z$ direction, which is the axial direction of the nanowire. We take both the values of these parameters and the chemical potential shift ($\Delta = 2.5$ meV) from Ref. \cite{Calavalle2022} (see also Ref. \cite{Suppmat}). We first calculate the spin texture at $E=-0.02$ eV, which is the Fermi level position found in \cite{Calavalle2022}, and plot it in the inset of Fig. \ref{fig_fig3}(a). Clearly, it shows that spins point normally to the Fermi surface, indicating that momentum (and velocity) is parallel to the spin, and therefore, one should expect a finite spin current according to Eq. \eqref{eq_Js2}. We proceed by numerically computing the spin current and plot its maximum value as a function of energy in Fig. \ref{fig_fig3}(a). Despite the inherent anisotropies, all spin components show comparable magnitudes, and increase linearly with increasing energy.

The fact that tellurium can be grown in a nanowire shape is very advantageous for our spin current generation. In strong spin-orbit materials, as seen in Fig. \ref{fig_fig2}(b) for TIs, the spin accumulates only at the edges and quickly decays in the bulk due to spin relaxation. In a two-dimensional material, this poses quite a challenge to detect a signal, since a nanometer-sized probe would be required at the edges. In contrast, in tellurium nanowires, even if the spin relaxes rapidly in the bulk, now the material edge corresponds to the surface of the wire, as shown in the inset of Fig. \ref{fig_fig3}(b). This means that the spins accumulate in a much larger scale that should be accessible to the experimental probes mentioned above. Using available spin relaxation times and lengths \cite{Niu2020, Suppmat}, we plot the spin accumulation across the nanowire's diameter in Fig. \ref{fig_fig3}(b). 

\textit{Discussion.---} We have identified a mechanism to generate time-dependent spin currents from non-equilibrium charge densities by continuously tuning the chemical potential in systems with spin-momentum locking. Such a method could in principle generate spin currents without the need of using magnetic materials, optical methods or charge current, therefore allowing for an all-electrical generation of spins in the absence of Joule heating. Not only is this attractive for energy-efficient electronic and spintronic devices, but the simple setup needed for the spin current generation could be easily integrated in existing microelectronic technologies.



One key ingredient of this work is a suitable spin-momentum locking to make Eq. \eqref{eq_Js} finite. Thus, any system wherein the spin texture changes with momentum has the potential to show similar results. Fortunately, a theory describing the different types of spin textures has recently been formulated based on symmetry analysis together with an exhaustive material search \cite{Acosta2021, Tan2022}, thus offering a plethora of material candidates to explore our proposed mechanism. 

Finally, we point out that our theory is not restricted to non-magnetic systems with spin-orbit coupling. Noncollinear antiferromagnets also present momentum-dependent spin texture \cite{Zelezny2017, Smejkal2022}, which, given the appropriate symmetries, should be able to generate spin current as well. Furthermore, by replacing the spin expectation with the orbital angular momentum operator \cite{Kontani2009, Canonico2020, Cysne2022}, one could calculate orbital currents which could be of relevance in chiral materials given their recently predicted parallel orbital-momentum locking \cite{Yang2023} and systems showing the orbital Rashba-Edelstein effect \cite{ElHamdi2023}. 







\begin{acknowledgments}
M. V. is grateful to Sergio O. Valenzuela, Roland K. Kawakami, Josep Fontcuberta, Shireen Adenwalla, Andrew D. Kent, Tiancong Zhu and Hossein Taghinejad for useful discussions. Both authors were supported by the Center for Novel Pathways to Quantum Coherence in Materials, an Energy Frontier Research Center funded by the US Department of Energy, Office of Science, Basic Energy Sciences. J.E.M. acknowledges a Simons Investigatorship.
\end{acknowledgments}

%

\end{document}